\documentstyle[12pt]{article}
\input{amssym.def}
\input{amssym.tex}
\input{epsf}
\catcode`\@=11
\@addtoreset{equation}{section}

\def\theequation{\thesection.\arabic{equation}}

\newtoks\@stequation

\def\subequations{\refstepcounter{equation}%
  \edef\@savedequation{\the\c@equation}%
  \@stequation=\expandafter{\theequation}
  \edef\@savedtheequation{\the\@stequation}
  \edef\oldtheequation{\theequation}%
  \setcounter{equation}{0}%
  \def\theequation{\oldtheequation\alph{equation}}}

\def\endsubequations{\setcounter{equation}{\@savedequation}%
  \@stequation=\expandafter{\@savedtheequation}%
  \edef\theequation{\the\@stequation}\global\@ignoretrue
  \vspace*{-12pt} \\}

\def\hybrid{\topmargin -20pt    \oddsidemargin 0pt
        \headheight 0pt \headsep 0pt
        \textwidth 6.25in       
        \textheight 9.5in       
        \marginparwidth .875in
        \parskip 5pt plus 1pt   \jot = 1.5ex}

\def\baselinestretch{1.2}

\catcode`\@=11

\def\marginnote#1{}
\newcount\hour
\newcount\minute
\newtoks\amorpm
\hour=\time\divide\hour by60
\minute=\time{\multiply\hour by60 \global\advance\minute by-\hour}
\edef\standardtime{{\ifnum\hour<12 \global\amorpm={am}%
        \else\global\amorpm={pm}\advance\hour by-12 \fi
        \ifnum\hour=0 \hour=12 \fi
        \number\hour:\ifnum\minute<10 0\fi\number\minute\the\amorpm}}
\edef\militarytime{\number\hour:\ifnum\minute<10 0\fi\number\minute}
\def\draftlabel#1{{\@bsphack\if@filesw {\let\thepage\relax
   \xdef\@gtempa{\write\@auxout{\string
      \newlabel{#1}{{\@currentlabel}{\thepage}}}}}\@gtempa
   \if@nobreak \ifvmode\nobreak\fi\fi\fi\@esphack}
        \gdef\@eqnlabel{#1}}
\def\@eqnlabel{}
\def\@vacuum{}
\def\draftmarginnote#1{\marginpar{\raggedright\scriptsize\tt#1}}

\def\draft{\oddsidemargin -.2truein
        \def\@oddfoot{\sl preliminary draft \hfil
        \rm\thepage\hfil\sl\today\quad\militarytime}
        \let\@evenfoot\@oddfoot \overfullrule 3pt
        \let\label=\draftlabel
        \let\marginnote=\draftmarginnote
   \def\@eqnnum{(\theequation)\rlap{\kern\marginparsep\tt\@eqnlabel}%
\global\let\@eqnlabel\@vacuum}  }

\def\preprint{\twocolumn\sloppy\flushbottom\parindent 2em
        \leftmargini 2em\leftmarginv .5em\leftmarginvi .5em
        \oddsidemargin -.5in    \evensidemargin -.5in
        \columnsep .4in \footheight 0pt
        \textwidth 10.in        \topmargin  -.4in
        \headheight 12pt \topskip .4in
        \textheight 6.9in \footskip 0pt
        \def\@oddhead{\thepage\hfil\addtocounter{page}{1}\thepage}
        \let\@evenhead\@oddhead \def\@oddfoot{} \def\@evenfoot{} }

\def\titlepage{\@restonecolfalse\if@twocolumn
\@restonecoltrue\onecolumn
     \else \newpage \fi \thispagestyle{empty}\c@page\z@
        \def\thefootnote{\fnsymbol{footnote}} }

\def\endtitlepage{\if@restonecol\twocolumn \else \newpage \fi
        \def\thefootnote{\arabic{footnote}}
        \setcounter{footnote}{0}}  

\catcode`@=12
\relax

\def\figcap{\section*{Figure Captions\markboth
        {FIGURECAPTIONS}{FIGURECAPTIONS}}\list
        {Figure \arabic{enumi}:\hfill}{\settowidth\labelwidth{Figure
999:}
        \leftmargin\labelwidth
        \advance\leftmargin\labelsep\usecounter{enumi}}}
 \relax
\def\tablecap{\section*{Table Captions\markboth
        {TABLECAPTIONS}{TABLECAPTIONS}}\list
        {Table \arabic{enumi}:\hfill}{\settowidth\labelwidth{Table
999:}
        \leftmargin\labelwidth
        \advance\leftmargin\labelsep\usecounter{enumi}}}
 \relax
\def\reflist{\section*{References\markboth
        {REFLIST}{REFLIST}}\list
        {[\arabic{enumi}]\hfill}{\settowidth\labelwidth{[999]}
        \leftmargin\labelwidth
        \advance\leftmargin\labelsep\usecounter{enumi}}}
 \relax

\font\fivesans=cmss10 at 4.61pt
\font\sevensans=cmss10 at 6.81pt
\font\tensans=cmss10
\newfam\sansfam
\textfont\sansfam=\tensans\scriptfont\sansfam=
\sevensans\scriptscriptfont
\sansfam=\fivesans

\mathchardef\endbar="375

\makeatletter
\newcounter{pubctr}
\def\publist{\@ifnextchar[{\@publist}{\@@publist}}
\def\@publist[#1]{\list
        {[\arabic{pubctr}]\hfill}{\settowidth\labelwidth{[999]}
        \leftmargin\labelwidth
        \advance\leftmargin\labelsep
        \@nmbrlisttrue\def\@listctr{pubctr}
        \setcounter{pubctr}{#1}\addtocounter{pubctr}{-1}}}
\def\@@publist{\list
        {[\arabic{pubctr}]\hfill}{\settowidth\labelwidth{[999]}
        \leftmargin\labelwidth
        \advance\leftmargin\labelsep
        \@nmbrlisttrue\def\@listctr{pubctr}}}
 \relax
\makeatother
\newskip\humongous \humongous=0pt plus 1000pt minus 1000pt
\def\caja{\mathsurround=0pt}
\def\eqalign#1{\,\vcenter{\openup1\jot \caja
        \ialign{\strut \hfil$\displaystyle{##}$&$
        \displaystyle{{}##}$\hfil\crcr#1\crcr}}\,}
\newif\ifdtup

\relax
\hybrid
\def\e{\epsilon}
\def\s{\sigma}
\def\m{\mu}
\def\n{\nu}
\def\t{\tau}

\def\cA{{\cal A}}
\def\limit#1#2{\smash { \mathop{#1} \limits_{#2} }  }
\def\be{\begin{equation}}
\def\ee{\end{equation}}
\def\pe{\quad . }

\font\ninerm=cmr9

\catcode`\@=11

\def\psfordvips{
\def\PSspeci@l##1##2{%
\d@my=0.1bp \d@mx=\drawingwd \divide\d@mx by\d@my%
\includegraphics{##1\space}}}

\newdimen\drawinght\newdimen\drawingwd
\newdimen\psxoffset\newdimen\psyoffset
\newbox\drawingBox

\newread\epsffilein    
\newif\ifepsffileok    
\newif\ifepsfbbfound   
\newif\ifepsfverbose   
\newdimen\epsfxsize    
\newdimen\epsfysize    
\newdimen\epsftsize    
\newdimen\epsfrsize    
\newdimen\epsftmp      
\newdimen\pspoints     
\def\ReadPSize#1{
\edef\PSfilename{#1}
\global\def\epsfllx{72}
\global\def\epsflly{72}
\global\def\epsfurx{540}
\global\def\epsfury{720}
\openin\epsffilein=#1
\ifeof\epsffilein\errmessage{I couldn't open #1, will ignore it}\else
   {\epsffileoktrue \chardef\other=12
    \def\do##1{\catcode`##1=\other}\dospecials \catcode`\ =10
    \loop
       \read\epsffilein to \epsffileline
       \ifeof\epsffilein\epsffileokfalse\else
          \expandafter\epsfaux\epsffileline:. \\%
       \fi
   \ifepsffileok\repeat
   \ifepsfbbfound\else
    \ifepsfverbose\message{No bounding box comment in #1;
using defaults}\fi\fi
   }\closein\epsffilein\fi
\def\psllx{\epsfllx}\def\pslly{\epsflly}%
\def\psurx{\epsfurx}\def\psury{\epsfury}%
\drawinght=\epsfury bp%
\advance\drawinght by-\epsflly bp%
\drawingwd=\epsfurx bp%
\advance\drawingwd by-\epsfllx bp%
}

{\catcode`\%=12
\global\let\epsfpercent=
\long\def\epsfaux#1#2:#3\\{\ifx#1\epsfpercent
   \def\testit{#2}\ifx\testit\epsfbblit
      \epsfgrab #3 @ @ @ \\%
      \epsffileokfalse
      \global\epsfbbfoundtrue
   \fi\else\ifx#1\par\else\epsffileokfalse\fi\fi}%

\def\epsfgrab #1 #2 #3 #4 #5\\{%
   \global\def\epsfllx{#1}\ifx\epsfllx\empty
      \epsfgrab #2 #3 #4 #5 @\\\else
   \global\def\epsflly{#2}%
   \global\def\epsfurx{#3}\global\def\epsfury{#4}\fi}%
\newcount\xscale \newcount\yscale \newdimen\pscm\pscm=1cm
\newdimen\d@mx \newdimen\d@my
\let\ps@nnotation=\relax
\def\psboxto(#1;#2)#3{\vbox{
   \catcode`\:=12
   \ReadPSize{#3}
   \divide\drawingwd by 1000
   \divide\drawinght by 1000
   \d@mx=#1
   \ifdim\d@mx=0pt\xscale=1000
         \else \xscale=\d@mx \divide \xscale by \drawingwd\fi
   \d@my=#2
   \ifdim\d@my=0pt\yscale=1000
         \else \yscale=\d@my \divide \yscale by \drawinght\fi
   \ifnum\yscale=1000
         \else\ifnum\xscale=1000\xscale=\yscale
                    \else\ifnum\yscale<\xscale\xscale=\yscale\fi
              \fi
   \fi
   \divide \psxoffset by 1000\multiply\psxoffset by \xscale
   \divide \psyoffset by 1000\multiply\psyoffset by \xscale
   \global\divide\pscm by 1000
   \global\multiply\pscm by\xscale
   \multiply\drawingwd by\xscale \multiply\drawinght by\xscale
   \ifdim\d@mx=0pt\d@mx=\drawingwd\fi
   \ifdim\d@my=0pt\d@my=\drawinght\fi
\message{[#3\space [BoundingBox\string:
\space\epsfllx\space\epsflly\space\epsfurx\space\epsfury]}%
\message{[scaled\space\the\xscale\string:
\space\the\drawingwd\space x \the\drawinght]]}%
 \hbox to\d@mx{\hss\vbox to\d@my{\vss
   \global\setbox\drawingBox=\hbox to 0pt{\kern\psxoffset\vbox to 0pt{
      \kern-\psyoffset
      \PSspeci@l{\PSfilename}{\the\xscale}
      \vss}\hss\ps@nnotation}
   \global\ht\drawingBox=\the\drawinght
   \global\wd\drawingBox=\the\drawingwd
   \baselineskip=0pt
   \copy\drawingBox
 \vss}\hss}
  \global\psxoffset=0pt
  \global\psyoffset=0pt
  \global\pscm=1cm
  \global\drawingwd=\drawingwd
  \global\drawinght=\drawinght
}}

\def\psboxscaled#1#2{\vbox{
  \catcode`\:=12
  \ReadPSize{#2}
  \xscale=#1
  \divide\drawingwd by 1000\multiply\drawingwd by\xscale
  \divide\drawinght by 1000\multiply\drawinght by\xscale
  \divide \psxoffset by 1000\multiply\psxoffset by \xscale
  \divide \psyoffset by 1000\multiply\psyoffset by \xscale
  \global\divide\pscm by 1000
  \global\multiply\pscm by\xscale
\message{[#2\space [BoundingBox\string:
\space\epsfllx\space\epsflly\space\epsfurx\space\epsfury]}%
\message{[scaled\space\the\xscale\string:
\space\the\drawingwd\space x \the\drawinght]]}%
  \global\setbox\drawingBox=\hbox to 0pt{\kern\psxoffset\vbox to 0pt{
     \kern-\psyoffset
     \PSspeci@l{\PSfilename}{\the\xscale}
     \vss}\hss\ps@nnotation}
  \global\ht\drawingBox=\the\drawinght
  \global\wd\drawingBox=\the\drawingwd
  \baselineskip=0pt
  \copy\drawingBox
  \global\psxoffset=0pt
  \global\psyoffset=0pt
  \global\pscm=1cm
  \global\drawingwd=\drawingwd
  \global\drawinght=\drawinght
}}

\def\psannotate#1#2{\def\ps@nnotation{#2\global\let
\ps@nnotation=\relax}#1}
\def\pscaption#1#2{\vbox{
   \setbox\drawingBox=#1
   \copy\drawingBox
   \vskip\baselineskip
   \vbox{\hsize=\wd\drawingBox\setbox0=\hbox{#2}
     \ifdim\wd0>\hsize
       \noindent\unhbox0\tolerance=5000
    \else\centerline{\box0}
    \fi
}}}

\def\at#1#2#3{\setbox0=\hbox{#3}\ht0=0pt\dp0=0pt
  \rlap{\kern#1\vbox to0pt{\kern-#2\box0\vss}}}

\newdimen\gridht \newdimen\gridwd
\def\gridfill(#1;#2){
  \setbox0=\hbox to 1\pscm
  {\vrule height1\pscm width.4pt\leaders\hrule\hfill}
  \gridht=#1
  \divide\gridht by \ht0
  \multiply\gridht by \ht0
  \advance \gridht by \ht0
  \gridwd=#2
  \divide\gridwd by \wd0
  \multiply\gridwd by \wd0
  \advance \gridwd by \wd0
  \vbox to \gridht{\leaders\hbox to\gridwd{\leaders\box0\hfill}\vfill}}

\catcode`\@=12 

\psfordvips

\jot = 1.5ex
\def\baselinestretch{1.65}
\parskip 5pt plus 1pt

\catcode`\@=11

\begin{document}
\overfullrule=0pt
\renewcommand{\theequation}{\thesection.\arabic{equation}}
\newcommand{\beq}{\begin{equation}}
\newcommand{\eeq}[1]{\label{#1}\end{equation}}
\newcommand{\ber}{\begin{eqnarray}}
\newcommand{\eer}[1]{\label{#1}\end{eqnarray}}

\begin{titlepage}
\begin{center}

\hfill CERN-TH/97-65\\
\hfill CPTH-S500-0497\\
\hfill hep-th/9707126\\

\vskip .5in

{\large \bf HETEROTIC / TYPE I DUALITY AND   D-BRANE INSTANTONS}
\vskip 1cm
{\bf  C. Bachas$\ ^{a}$, C. Fabre$\ ^a$, E. Kiritsis$\ ^b$,
 N.A. Obers$\ ^b$  and P. Vanhove$\ ^a$}
\vskip  0.5in
{$\ ^a$\  CPHT, Ecole Polytechnique,
 91128 Palaiseau, FRANCE\ \ \ \  \\
$\ ^b$  Theory Division, CERN, CH-1211, Geneva 23, SWITZERLAND\ \  }

\vskip .15in

\end{center}

\vskip .4in

\begin{center} {\bf ABSTRACT }
\end{center}
\begin{quotation}\noindent
\baselineskip 10pt

We study heterotic/type I
 duality in $d=8,9$ uncompactified dimensions.
We consider 
the  special (``BPS saturated'') ${\cal F}^4$ and
${\cal R}^4$ terms in the
effective one-loop heterotic action, which are expected to be
 non-perturbatively exact. Under the standard
duality map these translate to tree-level, perturbative  and
non-perturbative contributions on the type I side. We check 
agreement with the one-loop open string calculation,  and discuss
the  higher-order  perturbative
contributions, which arise because of the
mild non-holomorphicities of the heterotic elliptic genus. 
We put the heterotic world-sheet instanton corrections in a form
that can be motivated  as arising from a D-brane instanton
calculation on the type I side.

\end{quotation}
\vskip .2in
CERN-TH/97-65\\
July  1997\\

\end{titlepage}
\vfill
\eject
\def\baselinestretch{1.2}
\baselineskip 16 pt
\noindent
 \setcounter{equation}{0}


\section{Introduction}
\vskip 0.2cm

\hskip 0.4cm   The conjectured duality between the type I and heterotic SO(32)
string theories \cite{W1,PW}
occupies a special position in the web of dualities.
Together with the SL(2,{\bf Z}) symmetry of type IIB, it is  the only
duality that relates two  string theories in their 
critical dimension. It can thus be analyzed in flat space-time,
without the complications of  curved-compactification geometry.
Furthermore,  it is a duality between two
drastically different perturbative expansions.
In  heterotic theory  there is a single diagram of given genus, and
 ultraviolet divergences are cutoff by 
restricting the world-sheet  moduli to a fundamental domain.
In  type I theory, on the other hand, 
there are several   unoriented
surfaces with boundaries at any given order,
 and ultraviolet finiteness
results from  subtle cancellations  of their 
contributions \cite{Aug}. Finally, it can be argued \cite{Sen}  that the
heterotic/type I duality is the central piece of
the duality  web, from which all
other dualities 
can be derived  modulo  mild geometrical assumptions.

 One of the  aims of the present  paper  will be
 to strengthen the existing evidence
\cite{W1,PW,dab,Tseytlin,BaKi,ABFPT,FI}
for the equivalence of the type I and heterotic SO(32) theories.
We will in particular extend and sharpen the recent analysis
by two of us \cite{BaKi}  of 
special ${\cal F}^4$ and ${\cal R}^4$ terms of the effective
action in $d=8,9$  uncompactified
 dimensions\footnote{
The structure of
these  higher-derivative operators in theories with
sixteen supercharges has been of  interest recently
\cite{Hen,West,DS,DKMSW} for a  different reason: 
they are closely related to 
the velocity-dependent interactions of branes \cite{CB,Gut,Lif,DF,DKPS,
5d,rec,BB,long}, which are
being  analyzed vigorously in testing  the Matrix Theory
conjecture
 \cite{MM}.}. As we will argue   in the following section, 
there are good reasons to believe that the one-loop heterotic
calculation of these couplings is  {\it exact}. The only 
identifiable source
of  non-perturbative corrections are  
 heterotic five-brane instantons \cite{chs}.
These  need  a six-dimensional compact space to wrap around, 
and hence cannot  contribute to the effective action in $d>4$.
The situation on the type I side is on the other hand  different:
 first,  space-time 
supersymmetry  does not
{\it commute} with the genus expansion, so that
different terms of a superinvariant can be   generated  at
different orders \cite{Tseytlin}.
Secondly, in $d<9$ there are non-perturbative
corrections  from D1 instantons,  which are the duals of 
wrapped heterotic world-sheets \cite{PW,dab}.
Not surprisingly,  the one-loop heterotic calculation translates therefore 
under the standard duality map \cite{W1},
into a sum of tree-level, perturbative and
non-perturbative contributions.

  The pure-gauge one-loop  corrections on the type I side have been
computed previously in refs. \cite{BF,BaKi}. 
They are   given  by a  ten-dimensional super Yang-Mills expression  with a 
particular regularization of the (naively quadratically-divergent)
decompactification limit.
The same contribution  on the heterotic side comes as we will explain below
from  a sum
of infinite  towers of  BPS states, whose net effect is to  unfold  
the fundamental domain of the heterotic integral
into the strip. This  trick is well-known from
the study of finite-temperature partition functions \cite{McC,KS}. 
The decompactification limit 
 also agrees \cite{Tseytlin},  even though  it is regularized
 differently in the two expansions: the strip is replaced by 
a fundamental domain on the heterotic side, and by the  disk
and projective-plane diagrams on the type I side. It is a very interesting
question, whether an analogous geometric regularization exists
for the divergent loop of eleven-dimensional supergravity \cite{GGV}.

 In what concerns the  heterotic world-sheet instanton corrections,
we will put  them in this paper in a form that can be plausibly motivated 
 on the type I side. It is however an
open  (and we feel instructive)  problem, to learn how to calculate
these corrections directly from first principles. 
The logic can  in fact  be turned upside down: {\it assuming}
heterotic/type I duality, we can use the heterotic expression as
a guide to elucidate the rules of D-instanton calculus. 
These rules have been the subject of many interesting papers recently
\cite{Po,BBS,W,KSS,OV,HM,GG,B,GV,G}. All of them  involve inevitably
some guesswork,  since in contrast to conventional field theory
there is no functional-integral formulation at one's disposal.
One particular tricky point concerns   the correct counting of 
multiply-wrapped Euclidean branes \cite{BBS,OV,GG,GV}. Not surprisingly,
what we  find here is that one must include 
 all supersymmetric maps  of the D-string world-sheet onto the
compactification torus,
 modulo  (local and global) reparametrizations  of the former. 
This is of course  the heterotic world-sheet prescription,
which ensures  in particular  invariance under the O(d,d)
symmetry of space-time. It is tempting to 
conjecture that this is the correct  prescription 
in all instances, provided one extends reparametrization invariance
to include gauge transformations,  when gauge fields
live on the world volume of the brane.

This paper is organized as follows:
Section 2 describes the  one-loop 
heterotic calculation  
of  special  ${\cal F}^4$ and ${\cal R}^4$  terms,
  for vacua with  sixteen unbroken (real) supercharges, and its
relation to the (almost holomorphic) elliptic genus 
\cite{Schellekens,Lerche,Windey,ellwit}.
Section 3  reviews rather rapidly  how the one-loop type I calculation
of the pure-gauge ${\cal F}^4$ terms reduces to
 a (regularized) ten-dimensional super Yang-Mills expression \cite{BF}. 
In section 4 we employ  the unfolding trick to
 compare   the heterotic and type I results in
$d=9$ non-compact  dimensions. We also explain how the mild
non-holomorphicities of the elliptic genus translate to higher-order
perturbative corrections on the type I side.
In section 5  we move on to $d=8$, where world-sheet
instantons start to contribute.
 We express their  contribution in terms of the elliptic genus
of the complex structure  that is induced
from target space onto the string world-sheet.
This form can be motivated as arising from  an  instantonic
D-brane  calculation on the type I side, as we explain  in section 6.
Section 7   contains some concluding  remarks. 
A few  useful formulae are collected in the appendix.

\vskip 0.2cm
\section{One-Loop Heterotic  Thresholds}
\setcounter{equation}{0}

 The terms that will be of interest to us are those obtained
by dimensional reduction from the  ten-dimensional superinvariants, whose 
bosonic parts read \cite{roo,Tseytlin}
\be
\eqalign{
I_1=&t_8tr{\cal F}^4-{1\over4}\e_{10}Btr{\cal F}^4,\quad 
I_2=t_8(tr{\cal F}^2)^2-{1\over 4}\e_{10}B(tr{\cal F}^2)^2\cr
I_3=&t_8tr{\cal R}^4-{1\over 4}\e_{10}Btr{\cal R}^4 ,\quad
I_4=t_8(tr{\cal R}^2)^2-{1\over4}\e_{10}B(tr{\cal R}^2)^2\cr
I_5=&t_8(tr{\cal R}^2)(tr{\cal F}^2)-
{1\over4}\e_{10}B(tr{\cal R}^2)(tr{\cal F}^2)\pe\cr
}\ 
\label{1}
\ee
These are special because they contain anomaly-cancelling
CP-odd  pieces. As a result  anomaly cancellation fixes entirely their
coefficients in both the heterotic and the type I
 effective actions in ten dimensions. 
Comparing these coefficients
is not therefore  a test of duality, but rather of the fact that both
these theories are consistent \cite{Tseytlin}.
In lower dimensions things are different: the coefficients of the various
terms,  obtained from a single ten-dimensional superinvariant through dimensional
reduction,  depend on the compactification  moduli. 
Supersymmetry is expected to
relate  these coefficients  to each other, but is not   powerful
enough so as to fix them completely.
This is analogous
to the case of  N=1 super Yang-Mills in six dimensions:
the two-derivative
gauge-field  action is uniquely fixed, but after toroidal compactification
to four dimensions, it  depends on a holomorphic prepotential which
supersymmetry alone cannot determine.

 On the heterotic side there are good reasons to believe that these
dimensionally-reduced 
 terms receive only one-loop corrections. To start with, this is
true  for their  CP-odd anomaly-cancelling pieces 
\cite{ya}. 
Furthermore it has been
argued in the past \cite{cato}
that there exists a  prescription for treating  supermoduli, 
which  ensures that  space-time supersymmetry commutes with
the heterotic genus expansion, at least for vacua with more than
four  conserved supercharges\footnote{A  notable exception 
are compactifications with a naively-anomalous U(1) factor
\cite{Uone}.}.
Thus we may plausibly assume that
there are no higher-loop corrections to the terms of interest.
Furthermore, the only identifiable supersymmetric instantons
are the heterotic five-branes. These do not contribute in
  $d>4$ uncompactified dimensions, since they have  no finite-volume
6-cycle to  wrap around. Non-supersymmetric instantons, if they exist,
have  on the other hand  too many fermionic zero modes to
make  a non-zero contribution.  It should be noted that these
arguments do not apply to  the
 sixth superinvariant \cite{roo,Tseytlin}
\be
J_0= t_8 t_8 {\cal R}^4-{1\over 8}\e_{10}\e_{10}
{\cal R}^4 ,
\ee
which is not related to the anomaly.
 This receives as we will mention
below both perturbative and non-perturbative corrections.

\eject

The general form of the heterotic one-loop corrections to these
couplings is
\cite{Schellekens,Lerche}
\be
{\cal I}^{het}  =
-{\cal N}\
  \int_{ { F}}{d^2\tau \over
\tau_2^2}\; (2\pi^2 \tau_2)^{d/2} \   \Gamma_{d,d}\
{\cal A}({\cal F},{\cal R}, \tau)
\label{5}\ee
where ${\cal A}$ is an  (almost) holomorphic modular form
of weight zero
related to the elliptic genus, ${\cal F}$ and ${\cal R}$
 stand for the gauge-field
strength and curvature two-forms, $\Gamma_{d,d}$ is the lattice sum
over  momentum and winding modes
for  $d$ toroidally-compactified dimensions, ${ F}$ is the usual
fundamental domain,
 and 
\be
{\cal N} = {V^{(10-d)}\over 2^{10} \pi^6}
\ee
 is a
normalization that includes the volume of the uncompactified
dimensions \cite{BaKi}.
To keep things
 simple we have taken vanishing  Wilson lines on the
$d$-hypertorus, so that the sum over momenta ($p$) and windings ($w$),
\be
\Gamma_{d,d}  =
  \sum_{p,w}
 e^{-{\pi\tau_2\over 2} ( p^2 + w^2/\pi^2)
+ i \tau_1 p\cdot w} \ \ ,
\ee
factorizes inside the integrand.  Our conventions are
\be
 \alpha' = \frac{1}{2} \ ,\ \  q= e^{2\pi i\tau}\ ,
\ \  d^2\tau = d\tau_1 d\tau_2 \
\ee
while   winding  and   momentum  are normalized so that
$p\in {1\over L} {\Bbb Z}$ and
$w\in 2\pi L\ {\Bbb Z}$
for a circle of radius $L$.  The
Lagrangian form of the above lattice sum,
obtained by a Poisson resummation, reads
\be
\Gamma_{d,d}
 = \Bigl({2\over \tau_2}\Bigr)^{d/2}
\sqrt{\det G} \sum_{n_i,m_i \in {\Bbb Z}}
e^{-{2\pi\over\tau_2}\sum_{i,j}
 (G+B)_{ij} ( m_i \tau-n_i) ( m_j \bar\tau-n_j)}
\ee
with $G_{ij}$ the metric and $B_{ij}$ the (constant)
antisymmetric-tensor background
on the compactification torus. For a circle of radius $L$ the metric
is $G= L^2$.

The  modular function   ${\cal A}$ inside the integrand
  depends on the
 vacuum. It is {\it quartic, quadratic} or {\it linear} in ${\cal F}$ and
 ${\cal R}$,
for vacua with {\it maximal, half} or a {\it quarter}
 of unbroken supersymmetries.
The corresponding amplitudes have the property of saturating
exactly the fermionic zero modes in a Green-Schwarz light-cone
formalism, so that the contribution from left-moving oscillators
cancels out \cite{Lerche}\footnote{Modulo the regularization, ${\cal A}$
is in fact the appropriate term in the weak-field expansion of the
elliptic genus \cite{ell,Windey,ellwit}}.
 In the covariant NSR formulation this same
fact follows from $\vartheta$-function
 identities. 
As a result ${\cal A}$ should
have been
holomorphic in   $q$, but the use of a   modular-invariant
regulator  introduces some
extra  $\tau_2$-dependence \cite{Lerche}.
As a result ${\cal A}$ takes the  generic form 
of a finite polynomial in $1/\tau_2$,  with coefficients that have
Laurent expansions with at most simple poles in $q$,
\be
{\cal A}({\cal F},{\cal R},\tau) =  \sum_{r=0}^{r_{max}}\
 \sum_{n=-1}^\infty  {1\over\tau_2^r}\; q^n \
{\cal
A}^{(r)}_n({\cal F},{\cal R})
. \label{exp}
\ee
The poles in $q$ come from the would-be tachyon. Since this is not
charged under the gauge group, the poles are only present in the purely
gravitational terms of the effective action. This can be verified
explicitly in eq. (\ref{genus}) below.
The $1/\tau_2^r$ terms  play an important role in what follows.
They come from corners of the moduli space where vertex operators,
whose fusion can produce a massless state, collide. Each pair of
colliding operators  contributes  one factor of $1/\tau_2$.
For maximally-supersymmetric vacua the effective action of interest
starts with terms having four external legs, so that $r_{max} = 2$.
For vacua respecting  half the supersymmetries (N=1 in six dimensions
or N=2 in four) the one-loop effective action starts with terms
having two external legs and thus $r_{max} = 1$.

Much of what we will say in the sequel depends only on the above
generic properties of ${\cal A}$.  It will apply  in particular
in  the most-often-studied case of  four-dimensional
vacua with N=2.  For definiteness we will, however, focus
our attention to
the toroidally-compactified SO(32) theory, for which
 \cite{Schellekens,Lerche}
\be
\eqalign{
\ \ \  \cA( {\cal F},{\cal R},\tau )= &\  t_8\;  tr{\cal F}^4
\;+\;\frac{1}{2^7\cdot 3^2\cdot 5} \   {E_4^3\over \eta^{24}}\  t_8\;
tr{\cal R}^4
\;+\; {1\over 2^9\cdot 3^2} {\hat E^2_2 E_4^2\over \eta^{24}}
t_8\; ( tr{\cal R}^2)^2
 \cr &
+{1\over 2^9\cdot 3^2}\; \Bigl[  {E_4^3\over \eta^{24}} +
 {\hat E^2_2 E_4^2\over \eta^{24}}
-2 {\hat E_2E_4E_6\over \eta^{24}} -2^7\cdot
3^2\Bigr]\   t_8\; ( tr {\cal F}^2)^2 \cr \ \ \ \ \ &
+ {1\over 2^8\cdot 3^2}\; \Bigl[   {\hat E_2E_4E_6\over \eta^{24}}
 - {\hat E^2_2 E_4^2\over \eta^{24} }
\Bigr]\   t_8\;   tr {\cal F}^2   tr {\cal R}^2  \pe \cr} \
\label{genus}
\ee
Here $t_8$ is the well-known tensor appearing in four-point amplitudes
of the heterotic string \cite{GSW},
and  $E_{2k}$ are  the
Eisenstein series which are (holomorphic for $k>1$)
 modular forms of weight $2k$. Their explicit expressions are
 collected for convenience in the appendix.
The  second Eisenstein series $\hat E_2$ is special, in that it
 requires non-holomorphic regularization.
 The entire non-holomorphicity of ${\cal A}$ in  eq. (\ref{genus}),
arises through  this modified Eisenstein series.

In the toroidally-compactified  heterotic string
all one-loop  amplitudes with fewer
than four external legs vanish identically \cite{AS}.
Consequently  eq. (\ref{5}) gives directly the effective action,
without
the need to subtract  one-particle-reducible diagrams,  as is the case
at tree level \cite{slo}. Notice also that this four-derivative
effective action has infrared divergences  when more than one
dimensions are
compactified.  Such IR divergences can be regularized in
a modular-invariant way with
a curved background \cite{KK,chem}. This should be kept in mind,
 even though for the sake of simplicity we will be working
in this paper  with
unregularized expressions.

\vskip 0.8cm

\section{One-loop Type-I Thresholds}
\vskip 0.2cm
\setcounter{equation}{0}


 The one-loop type I effective action has the form
\be
{\cal I}^{I}  = -{i\over 2}({\cal T} + {\cal K} + {\cal A} + {\cal M})
\ee
corresponding to the contributions of the torus, Klein bottle,
annulus and M{\"o}bius strip diagrams. Only the last two surfaces
(with boundaries) contribute to the ${\cal F}^4$, $({\cal F}^2)^2$
and ${\cal F}^2{\cal R}^2$ 
terms of the action. The remaining two pure gravitational
terms may also receive  contributions from the torus and from the
Klein bottle. Contrary to what happens on the heterotic side,
this one-loop calculation is corrected by  both
higher-order perturbative and non-perturbative contributions.

  For the sake of completeness we review here the calculation of pure
gauge terms following refs. \cite{BF,BaKi}. To the order of
interest only the short BPS multiplets of the open string spectrum
contribute. This follows from the fact that the wave operator
in the presence of a background magnetic field ${\cal F}_{12}=
{\cal B}$  reads
\be
{\cal O} = M^2+ (p_{\perp})^2 + (2n+1)\epsilon + 2\lambda \epsilon
\ee
where $\epsilon \simeq {\cal B} + o({\cal B}^3)$ is a non-linear
function of the field, $\lambda$ is the spin operator projected
onto the plane (12),  $p_{\perp}$
denotes the momenta in the  directions $034\cdots  9$,
 $M$ is a string mass
and $n$ labels the Landau levels. The one-loop free energy thus formally
reads
\be
{\cal I}^{I} = - {1\over 2}
 \int_0^\infty
{dt\over t}\ 
Str\   e^{-{\pi
t\over 2}{\cal O}} 
\ee
where the supertrace stands for a sum over all bosonic minus
fermionic states of the open string, including a sum over 
the Chan-Paton charges, the center of mass
positions and momenta, as well as over the Landau levels.

Let us concentrate on the spin-dependent term 
inside the integrand, which can be expanded for weak field
\be
e^{-\pi t  \lambda\epsilon }= \sum_{n=0}^\infty
 {\left(-\pi t \right)^n\over n!}\left(\lambda\epsilon\right)^n \pe
\ee
The $n<4$ terms vanish for every supermultiplet because of the
properties of the helicity supertrace \cite{BaKi},
while  to the $n=4$ term  only short BPS
multiplets can contribute. The only short multiplets in  the perturbative
spectrum of the  toroidally-compactified
 open string are the SO(32)  gauge bosons and their
Kaluza-Klein dependents. It follows after some straightforward algebra
 that the special ${\cal F}^4$ terms of interest are given by
the following (formal) one-loop super Yang-Mills expression 
\be
{\cal I}^{I} = -{V^{(10-d)}\over
 3\cdot 2^{12}\pi^4}\  
 \int_0^\infty {dt\over t} (2\pi^2 t)^{{d\over 2}-1}
\sum_{p\in {}^*\Gamma}  e^{-\pi t p^2/2}\  \times\  t_8 {\rm Tr}_{adj}
 {\cal F}^4
\ee
where ${}^*\Gamma$ is the lattice of Kaluza-Klein momenta on
a $d$-dimensional torus, and the trace is in the adjoint representation
of SO(32).

 This expression is quadratically UV  divergent, but
in the full string theory one must remember to (a) regularize contributions
from the annulus and M{\"o}bius uniformly in the {\it transverse}
closed-string  channel, and (b) to subtract the one-particle-reducible
diagram corresponding to the exchange of a massless (super)graviton
between two $tr{\cal F}^2$ tadpoles, with the trace being here
in the fundamental representation of the group.
 The net result can be summarized easily, after a Poisson resummation
from the open-channel Kaluza-Klein momenta to the closed-channel
windings, and  amounts to simply subtracting the contribution of the
zero-winding sector 
\cite{BF,BaKi}. 
 Using also the fact
that ${\rm Tr}_{adj}
 {\cal F}^4 = 24 tr {\cal F}^4 + 3 (tr {\cal F}^2)^2$ we thus derive
the final one-loop expression on the type I side
\be
{\cal I}^{I} = -{V^{(10)}\over
  2^{10}\pi^6}\  
 \int_0^\infty {dt\over t^2} 
\sum_{w\in \Gamma\backslash\{0\}} 
 e^{-  w^2/2\pi t}\  \times\  t_8 \Bigl( tr {\cal F}^4 + {1\over 8}
 (tr {\cal F}^2)^2 \Bigr)
 \ .
\ee
The conventions for momentum and winding are the same as in
the heterotic calculation of the previous section.

  The calculation of the gravitational terms is more involved
because we have no simple background-field method at our disposal.
It can be done in principle
 following the method described in ref. \cite{ABFPT}. There is one
particular point we want to stress here: 
if the  one-loop heterotic calculation is exact, and assuming that
duality is valid, there should be no world-sheet instanton corrections
on the type I side. Such corrections would indeed translate to
non-perturbative contributions  in the heterotic string
\cite{silver}, and we have
just argued above that there should not  be any. The dangerous diagram
 is the torus which can wrap non-trivially around the compactification
manifold.
The type I torus diagram  is on the other hand
identical to the type IIB one,  assuming there are  only graviton insertions.
This latter diagram was  explicitly calculated  
in eight uncompactified dimensions  in ref. \cite{Boris},  confirming 
our  expectations: the CP-odd invariants
  only  depend  on the complex
structure of the compactification torus, but not  on
its K{\"a}hler structure. This is not true for the CP-even invariant
$J_0$.

\vskip 0.8cm

\section{Circle  Compactification}
\vskip 0.2cm
\setcounter{equation}{0}

Let us  begin now our comparison  of the effective actions  with the
simplest situation, namely
  compactification on a circle. There are no world-sheet or D-string
instanton contributions 
in this case, since Euclidean world-sheets have no
finite-area
manifold in target space
to wrap around.  Thus the one-loop heterotic
amplitude  should be expected to match
with a perturbative calculation on the type I side.
This sounds at first puzzling,  since the heterotic theory  contains
infinitely more charged BPS multiplets than the type I theory in its
perturbative spectrum. Indeed, one can combine
any state of the $SO(32)$ current
algebra  with appropriate $S^1$-winding and momentum,
so as to satisfy   the level-matching condition of physical states.
The heterotic theory thus contains short multiplets
in arbitrary representations of the gauge group.

The puzzle is resolved by  a well-known trick, used 
previously in the study of string thermodynamics \cite{McC,KS}, and which  
 trades  the winding sum for an unfolding of the
fundamental domain  into the half-strip,
$-{1\over 2} < \tau_1 < {1\over 2}$ and $\tau_2>0$.
The trick works  as follows: starting with
the Lagrangian form of the heterotic lattice sum,
 \be
(2\pi^2\tau_2)^{1/2}\  \Gamma_{1,1}  = 2\pi L
  \sum_{(m,n) \in {\Bbb{Z}}^2 }
 e^{- 2\pi L^2\vert m\tau -n\vert^2/\tau_2}
\pe \ee
one decomposes  any non-zero pair of  integers   as
$(m,n) = (jc,-jd)$, where $j$ is their greatest common divisor
(up to a sign). We will denote the set of all relative primes
$(c,d)$, modulo an overall sign, by 
${\cal S}$. 
The lattice sum can thus  be written as
\be
({2\pi^2\tau_2})^{1/2} \  \Gamma_{1,1}
 = 2\pi L \Bigl[\  1+
 \sum_{j\in  {\Bbb{Z}}\backslash \{0\}}\   \sum_{(c,d)\in {\cal S} }
 e^{- 2\pi L^2 j^2\vert c\tau +d \vert^2/\tau_2} \Bigr] \pe
\ee

Now the set ${\cal S}$ is in one-to-one correspondence with
all modular
transformations,
\be
{\tilde \tau} = {a\tau +b\over c\tau+d}\ \  \Longrightarrow\
{\tilde\tau}_2 = {\tau_2\over \vert c\tau+d\vert^2} \
\ee
such that
$
-{1\over 2} < {\tilde\tau}_1 \le {1\over 2}\ $.
 Indeed  the condition  $ad-bc=1$ has a solution  only if
 $(c,d)$ belongs  to  ${\cal S}$,
and the solution is unique modulo a shift and an irrelevant sign
\be
\left( \matrix{a  & b \cr c & d}\right)
 \rightarrow \pm  \left( \matrix{1 &{ l} \cr 0 & 1}\right)
\left( \matrix{a  & b \cr c & d}\right)\ .
\label{transfo}
\ee
By choosing  ${ l}$  appropriately we may always bring $\tilde\tau$
inside the strip, which establishes the above  claim.

Using the modular invariance of ${\cal A}$, we can thus suppress the
sum
over $(c,d)\in \cal S$ and unfold  the integration regime
for the $j\not= 0$ part of the expression.  This gives
\be
{\cal I}^{het} =
 -{V^{(9)} L\over 2^9 \pi^5}\; 
\Biggl[  \int_{{ F}}{d^2\tau \over
\tau_2^2}\;{\cal A}
\ + \
\int_{\rm strip} {d^2\tau\over\tau_2^{\ 2}}\;
 \sum_{j \neq 0}
 e^{- 2\pi L^2
j^2/\tau_2 }\;{\cal A}\; \Biggr]
   \pe
\label{trick}
\ee
There is  one subtle point in this  derivation
\cite{KS}:
convergence of the original threshold integral, 
 when ${\cal A}$ has a
 ${1\over q}$ pole\footnote{(Physical) massless states do not lead to
IR divergences in four-derivative  operators in nine dimensions},
 requires that we integrate 
 $\tau_1$ first in the  $\tau_2\to\infty$ region.
Since constant $\tau_2$ lines
transform  however non-trivially under SL(2,{\bf Z}),
the integration over the entire strip would have to be supplemented
by a highly  singular  prescription.
The problem could be  avoided
 if  integration of the $m\not= 0$ terms in the Lagrangian
 sum (i.e. those  terms that required a change of integration variable)
 were  absolutely convergent. This is the case for $L>1$,
so 
 expression (\ref{trick}) should only be trusted in
this  region.

\begin{figure}
\begin{center}
\leavevmode
\epsfxsize=16cm

\epsfbox{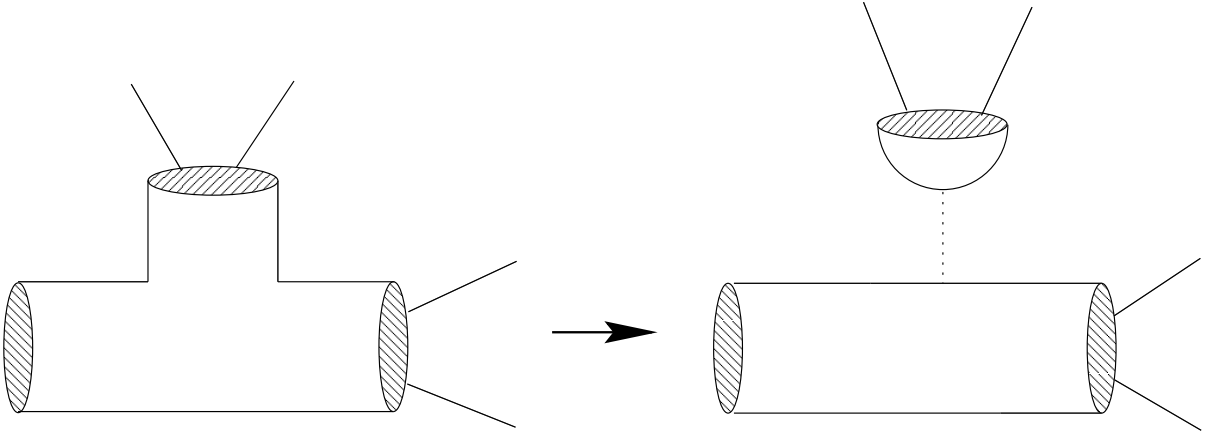}
\end{center}
\caption[]{\ninerm
A type I diagram with Euler characteristic
$\chi=-1$. This contributes to the
  $(tr{\cal F}^2)^2$ piece of the
effective action,  only in degeneration limits such as the one depicted above.}
\label{f1}\end{figure}

 Let us now proceed to evaluate this  expression. The fundamental
domain
integrals can be performed explicitly by using the formula
\cite{Lerche}
\be
 \int_{ F} {d^2\tau\over \tau_2^2}
 ({\hat E}_2)^r  \Phi_{r} =
\  {\pi\over 3(r+1)} [c_0 -24(r+1) c_{-1}]
\ee
where
\be
\Phi_{r}(q) =  \sum_{n=-1}^\infty c_n q^n
\ee
is any modular form of weight $-2r$ which is holomorphic everywhere
except possibly for a simple pole at zero.
As for the strip integration, it picks up only the ${\cal O}(q^0)$
term in the expansion of ${\cal A}$.
Modulo  the non-holomorphic regularization, 
only the SO(32)  gauge bosons contribute to the elliptic
genus at this order, in agreement precisely with the result of  the
type I side!
 For $k\ge 1$ let us  define more generally
\be
 \; \int_0^\infty {d\tau_2\over \tau_2^{1+k}} \sum_{j\not= 0}
 e^{-2\pi L^2 j^2/\tau_2} =  {2 \Gamma(k) \zeta(2k)
 \over (2\pi L^2)^k}\ \equiv {N_k\over L^{2k}},
\label{Zag}
\ee
where $L$ is the radius of the compactification circle.
The one-loop SO(32) heterotic action takes finally the form
\be
\eqalign{
 {\cal I}^{het} =  -{ V^{(10)} \over 2^{10}\pi^6}&\;
\Biggl\{ \;\; 
  {\pi\over 3}  \Bigl[
  \; {\cal F}^4\; - {1\over 8}\; {\cal F}^2 {\cal R}^2
  + {1\over 8}\; {\cal R}^4 \;
+\; {1\over 32} \left(  {\cal R}^2  \right)^2  \Bigr] + \cr
+  & {N_1\over L^2} 
 \Bigl[
   \; {\cal F}^4\; + {1\over 8}  ( {\cal F}^2)^2\;
-{5\over 16} \;  {\cal F}^2 {\cal R}^2  +  {31\over
240}  {\cal R}^4 \;
+\frac{19}{ 192 }\;  \left(  {\cal R}^2  \right)^2  \Bigr] - \cr
-{5\over 16\pi}\times  &
 {N_2\over L^4} 
 \Bigl[
3  ( {\cal F}^2)^2\;
 - {5} \;  {\cal F}^2  {\cal R}^2  +\; 2
  \left(  {\cal R}^2  \right)^2  \Bigr]
 \ \;  +\;\  {21 \over 64\pi^2}\times  
 {N_3\over L^6}
 ( {\cal F}^2  -  {\cal R}^2)^2\;  \Biggr\} \pe}
\label{s1}
\ee
To simplify notation we have written here ${\cal F}^4$ instead of
 $t_8\; tr {\cal F}^4$,
  $({\cal F}^2)^2$ instead of  $t_8\;  tr {\cal F}^2 tr {\cal F}^2$
etc.

We have
 expressed the result as an expansion in inverse
powers of the compactification volume.
Since the heterotic/type I duality map
transforms ($\sigma$-model) length scales as
\be
L^2_h = L^2_I/\lambda_I
\ee
with $\lambda_I$ the open-string loop counting parameter, this
expansion
can be translated to a genus expansion on the type I side.
The Euler number
 of an non-orientable surface is given by
$\chi=2-2g-B-C$
where $g$ is the number of holes, $B$ the number of boundaries
and $C$ the number of cross-caps.
The leading term corresponds to the disk and projective plane 
diagrams
and is completely fixed
by ten-dimensional
supersymmetry and  anomaly cancellation \cite{Tseytlin}.
To check this one must remember to transform
the metric in  both $V^{(10)}$ and the tensor $t_8$ appropriately.
Notice that the type I sphere diagram, which is the same as in type IIB,
only contributes to the $J_0$ invariant which we are not considering
here.
The subleading o($L^{-2}$) terms correspond to the annulus, M{\"o}bius
strip, Klein bottle and torus diagrams, all with $\chi = 0$. 
For zero background
curvature these agree  with the type I calculation \cite{BaKi}
as described  in section 3.

  The last two  terms in the expansion (\ref{s1}) correspond 
to diagrams with $\chi = -1,-2$. These contributions must be there
if the  duality map of ref. \cite{W1} does not receive higher-order
corrections. Such corrections could anyway always be absorbed
by redefining fields on the type I side, so that if duality holds,
there must exist  some regularization scheme in which these higher-genus
contributions do arise. These terms do on the other hand come from
the boundary of moduli space. For instance the $\chi =-1$ contribution
to the $({\cal F}^2)^2$ term comes from the boundary of moduli
space shown in figure 1. It could thus be conceivably 
eliminated in favour of some lower-dimension operators in the effective
action.  
 
    It is in any case striking that a single heterotic diagram contains
contributions from different topologies on the type I side. 
Notice in particular that the divergent $w=0$ term in the one-loop
field theoretic calculation,  regularized on the heterotic side
by replacing  the strip by a fundamental domain, is regularized on
the type I side by replacing the annulus by the disk.


\section{ Two-torus Compactification}

\vskip 0.2cm

 The next simplest situation corresponds to compactification on a
two-dimensional torus. There are in this case world-sheet
instanton contributions  on the heterotic side, and our aim in
this and the following sections will be to understand them as
(Euclidean)
D-string trajectory  contributions on the type I side.
 The discussion can
be extended with little effort to toroidal compactifications in
lower than eight dimensions. New effects are only expected to arise
in  four or fewer  uncompactified dimensions, where the solitonic
heterotic instantons  or,  equivalently, 
 the type I  D5-branes  can   contribute.

 The target-space torus 
 is  characterized by  two complex moduli,
the  K{\"a}hler-class
\be
 T = T_1 + iT_2 = \frac{1}{\alpha' } (B_{89} + i\sqrt{G})
\ee
and the complex structure
\be
U = U_1 + i U_2 = ( G_{89} + i\sqrt{G} )/ G_{88} \ ,
\ee
where $G_{\mu\nu}$ and $B_{\mu\nu}$ are the
 $\sigma$-model metric and antisymmetric
tensor  on the heterotic side.
The one-loop  thresholds now read
\be
{\cal I}^{ het}=
{V^{(8)}\over 2^9 \pi^4} \; \int_{ F}{d^2\tau\over
  \tau_2}\Gamma_{2,2}\;  {\cal A}({\cal F},{\cal R},\tau) \ , 
\label{dd8}\ee
where the lattice sum takes the form  \cite{DKL}
\be
    {\Gamma_{2,2}}  =
{ T_2\over \tau_2 }\times
 \sum_{ M \in {\rm Mat}(2 \times 2, {\Bbb{Z}}) }
 e^{ 2\pi i  T {\rm det}M }
e^{- \frac{\pi T_2 }{ \tau_2 U_2 }
\big| (1\; U)M  \big( {\tau \atop -1} \big) \big| ^2 } \pe
\label{DKL}
\ee
The exponent  in the above sum is (minus) the Polyakov action,
\be
S_{\rm Polyakov} = {1\over 4\pi\alpha^\prime} \int d^2 \sigma ( \sqrt{g}
  G_{\mu\nu}g^{\alpha\beta}\partial_\alpha X^\mu \partial_\beta X^\nu
+i  B_{\mu\nu}\epsilon^{\alpha\beta}
\partial_\alpha X^\mu \partial_\beta X^\nu )\ ,
\ee
evaluated 
for the topologically non-trivial mapping of the string world-sheet
onto the target-space torus,
\be
\left(\matrix{X^8\cr X^9\cr}\right)=
 M \left(\matrix{\s^1\cr \s^2\cr}\right)\equiv
\left(\matrix{m_1&n_1\cr
m_2&n_2\cr}\right)\left(\matrix{\s^1\cr \s^2\cr}\right)
 \  \pe
\label{wrap}
\ee
The entries of the matrix $M$ are  integers, and both target-space
and world-sheet coordinates take values in the (periodic) interval
$(0, 2\pi]$. To verify the above assertion one needs to use
the metrics
\be
 G_{\mu\nu} = {\alpha^\prime T_2\over U_2}\; 
\left(\matrix{ 1 & U_1 \cr U_1 & \vert U\vert^2 \cr}
\right)  \; ,
\ \ \ 
g^{\alpha\beta} = {1\over \tau_2^{\ 2}
}\;  \left(\matrix{ \vert\tau\vert^2 & -\tau_1 \cr
-\tau_1 & 1 \cr}\right) \ . 
\ee

The Polyakov action is invariant under global reparametrizations
of the world-sheet,
\be
\left( \matrix{\sigma^1 \cr \sigma^2 \cr}\right)
\rightarrow
\left( \matrix{a & -b\cr -c & d\cr}\right)
\left( \matrix{\sigma^1 \cr \sigma^2 \cr}\right)\ ,
\ee
which transform
\be
\tau \rightarrow    {a\tau+b\over c\tau +d}\ , \ \ {\rm and}\ \ 
M \rightarrow  
 M \ \left( \matrix{d & b\cr c & a\cr}\right) \ 
\pe
\ee
Following Dixon, Kaplunovsky and Louis \cite{DKL}, we
decompose the set of all matrices $M$  into orbits of PSL(2,{\bf Z}),
which is the group of the above transformations up to an overall sign.
 There are three types of orbits,
$$
\eqalign{& {\rm invariant}:  \ M=0 \cr
&  {\rm degenerate}: \  {\rm det} M = 0 ,\;  M\not= 0   \cr
&  {\rm non-degenerate}: \  {\rm det} M \not= 0  \cr}
$$
A  canonical choice of representatives for the  degenerate orbits
is
\be
M  = \left( \matrix{0  & j_1\cr 0  & j_2 \cr}\right)
\ee
where the
 integers $j_1,j_2$ should not both vanish,
but are  otherwise arbitrary.
Distinct elements of  a degenerate orbit are in one-to-one correspondence
with the set ${\cal S}$, i.e.
with modular transformations that map the fundamental domain inside
the strip, as in section 4.
In what concerns the  non-degenerate
orbits, a canonical choice of representatives is
\be
 M = \pm \left(\matrix{ k& j\cr 0&p\cr}\right) \ \ 
{\rm with}\ \   0\le j <k \quad, \quad \ p\not= 0\ .
\label{nondeg}
\ee 
Distinct elements of a non-degenerate orbit
 are  in one-to-one correspondence with the fundamental
domains of $\tau$ in the upper-half  complex plane.

Trading  the sum
over orbit elements for  an extension of the integration region
of $\tau$, we can thus express eqs. (\ref{dd8},\ref{DKL}) as
follows  

\be
\eqalign{
{\cal I}^{het} =
-{V^{(8)}  T_2 \over 2^9 \pi^4}  \times &
\Biggl\{  \int_{F}{d^2\tau \over
\tau_2^2}  {\cal A} 
\ + \
\int_{\rm strip} {d^2\tau\over\tau_2^{\ 2}}
\sum_{(j_1,j_2) \neq (0,0)}
 e^{- \frac{\pi T_2 }{ \tau_2 U_2 }
\big| j_1+j_2U \big| ^2 }
\;{\cal A} \cr
&+
2\; \int_{\Bbb C^+} {d^2\tau\over\tau_2^{\ 2}}
 \sum_{{0 \leq j<k} \atop { p\neq 0}}
        e^{2\pi i Tpk}
        \; e^{- \frac{\pi T_2 }{ \tau_2 U_2 }
        \big|k\tau - j-pU \big| ^2 }\; {\cal A}
\Biggr\} \ \equiv {\cal I}_{pert} +{\cal I}_{inst} .
\cr}
\label{3terms}
\ee
The three terms inside the curly brackets are constant,
power-\discretionary{}{}{}sup\-pres\-sed
and exponen\-tially-\discretionary{}{}{}suppressed in the large
compactification-volume 
limit. They correspond to tree-level, higher perturbative and
 non-perturbative, respectively, contributions
on the type I side. The discussion of the perturbative contributions
follows exactly the analogous discussion in section 4. The only
difference is the replacement of eq. (\ref{Zag}) by
\be
\eqalign{
\int_0^\infty {d\tau_2\over\tau_2^{\ 1+k}}\;\sum_{(j_1,j_2) \neq (0,0)}
 e^{- \frac{\pi T_2 }{ \tau_2 U_2 }
\big| j_1+j_2U \big| ^2 }
=&
\Gamma(k) \left({U_2\over\pi T_2}\right)^k \; 
\sum_{(j_1,j_2) \neq (0,0)} \vert j_1 + j_2 U\vert^{-2k}\cr
=& {2 \Gamma(k) \zeta(2k)\over (\pi T_2)^k} E(U,k) . \cr}
\ee
where $E(U,k)$ are generalized Eisenstein
 series \cite{Zagier}.
In the open-string channel of the type I side this takes into
account properly the (double) sum over Kaluza-Klein momenta 
\cite{BaKi}. Notice that the holomorphic anomalies  in ${\cal A}$ 
lead again to higher powers of the inverse volume, which translate
to higher-genus contributions on the type I side. Notice also that
the $k=1$ term has a logarithmic infrared divergence, which must be
regularized appropriately, as discussed in the introduction.

 We turn now to the  novel feature of eight dimensions, namely
the contributions  of world-sheet instantons. 
Plugging in the expansion (\ref{exp}) of the elliptic genus, we
are lead to consider the integrals
\be
I_{n,r} = 
\int_{\Bbb C^+} {d^2\tau\over\tau_2^{\ 2}}
         \; e^{- \frac{\pi T_2 }{ \tau_2 U_2 }
        \big|k\tau - j-pU \big| ^2 }\; {1\over \tau_2^{\ r}} e^{2i\pi\tau n}
\label{form2}
\ee
Doing first the (Gaussian) $\tau_1$ integral, 
one finds after some rearrangements
\be
I_{n,r} = {1\over k} \sqrt{U_2\over T_2} e^{2i\pi n({j+pU_1\over k})}
e^{2\pi kp T_2} \int_0^\infty
{d\tau_2\over \tau_2^{3/2+r}} e^{-{\pi T_2\over U_2}(k+{n U_2\over kT_2})^2
\tau_2}  e^{-\pi p^2 T_2 U_2/\tau_2}
\ee
The $\tau_2$ integration can now be done using the formula
\be
\int_0^\infty {dx\over x^{3/2+r}}  e^{-ax-b/x}
= \left(- {\partial\over \partial b}\right)^r \sqrt{\pi\over b}\;
 e^{-2\sqrt{ab}} 
\label{formula}
\ee
where  $a ={\pi T_2\over U_2}(k+{n U_2\over kT_2})^2$ and 
$b= \pi p^2 T_2 U_2$ are both proportional to the volume of the
compactification  torus.  
The leading term in the large-volume limit  is obtained when all
derivatives hit  the exponential in the above expression.
Using (\ref{formula}) we find
\be
I_{n,r} = {1\over k\vert p\vert T_2} \left( {k\over
 \vert p\vert U_2} \right)^r \; e^{2\pi k (p-\vert p\vert)T_2}
e^{2i\pi n [ {j+pU_1\over k} + i\vert p\vert {U_2\over k} ]}
\; \Bigl( 1 + {\rm o}({1\over T_2})\Bigr)
\label{formula1}
\ee
and plugging back into eq. ({\ref{3terms}) we get
\be
 {\cal I}^{het}_{inst} \simeq -{2 V^{(10)}\over 2^{10}\pi^6}
  \sum_{{0 \leq j<k} \atop { p > 0}}
      {1\over k p T_2}\;  e^{2\pi i Tpk}
        \; {\cal A}\left({j+p U\over k}\right) +\  {\rm c.c.} 
\label{inst}
\ee
This equality is exact for the holomorphic parts of the elliptic genus.
Correction terms have the form of an order-$r_{max}$ polynomial
in inverse powers of the volume, as we will discuss in a minute.

 Expression (\ref{inst}) has an elegant rewriting in terms of 
Hecke operators  $H_N$ \cite{DMVV}.
On  any modular form
 $\Phi_r(z)$ of weight $-2r$,
the action of a Hecke operator,  defined by  \cite{Serre}
\be
H_N[\Phi_r](z) = {1\over N^{2r+1}} 
\sum_{k,p>0\atop kp=N} \sum_{0\le j <k } k^{2r}\;
  \Phi_r\left(pz+j\over
k\right) \ ,
\label{hec}\ee
gives another modular form of the same  weight.
The Hecke operator is self-adjoint
with respect to the inner product defined by integration of modular
forms on a fundamental domain.
Using the above definition
 one finds
\be
{\cal I}^{het}_{inst} \simeq -{2 V^{(10)}\over 2^{10}\pi^6}  \sum_{N=1}^\infty
      {1\over  T_2}\;  e^{2\pi i NT}
        \; H_N[ {\cal A}](U) + {\rm c.c.} 
\label{inst1}
\ee 
In the above  form the result might be easier to compare with
a calculation based on the heterotic matrix string theory
 \cite{Matrixstring}.


  Let us complete now the calculation, by taking into account the
sub-leading terms in the large-volume limit. 
Using eq. ({\ref{formula}) we can in fact evaluate explicitly
the integrals ({\ref{form2}). After some long but straightforward
algebra the correction terms can all be expressed in terms
of the induced moduli
\be
{\cal U}= {j+pU\over k}\ \ \ {\rm and } \ \ \ 
{\cal T}= kp T \ .
\ee     

\be
I_{n,1} \to I_{n,1}\times \left(\; 1 + {1\over {\cal T}_2}( n 
 {\cal U}_2  + {1\over 2\pi}) \right)\ ,
\ee
\be
I_{n,2} \to I_{n,2}\times \left(\;1 +  {1\over {\cal T}_2}( 2n
 {\cal U}_2 +  {3\over 2\pi})
+ {1\over {\cal T}_2^2}( n^2 {\cal U}_2^2 + {3n {\cal U}_2\over 2\pi}
+ {3\over 4\pi^2}) \right) \ .
\ee
These terms can be rewritten elegantly by using the 
 operator
\be
\square  \equiv {\cal U}_2^2 \partial_{\cal U} {\bar \partial}_{\cal U}
\ee
This is a modular invariant operator, which annihilates
holomorphic forms. 
The correction terms for all $r=0,1,2$ are summarized 
by the expression
\be
{\cal U}_2^{\ r} e^{- 2i\pi {\cal U}n}
\left( 1 + {1\over \pi {\cal T}_2}\square + {1\over 2}
{1\over \pi^2 {\cal T}_2^2 } (\square^2 - \square/2)\right)\
{\cal U}_2^{\; -r}  e^{ 2i\pi {\cal U}n} .
\ee
The instanton sum is modified accordingly to
\be
{\cal I}^{het}_{inst} = -{2 V^{(10)}\over 2^{10}\pi^6}  \sum_{instantons}
      {1\over  {\cal T}_2}\;  e^{2\pi i {\cal T}}
        \; \left( 1 + {1\over \pi {\cal T}_2}\square + {1\over 2}
{1\over \pi^2 {\cal T}_2^2 } (\square^2 - \square/2)\right)\
 {\cal A}({\cal U})\  + \  {\rm c.c.} \ .
\label{inst3}
\ee
One final rearrangement puts this to the form
\be
{\cal I}^{het}_{inst} = - {2 V^{(10)}\over 2^{10}\pi^6}  \sum_{instantons}
      {1\over  {\cal T}_2}\;  e^{2\pi i {\cal T}}
        \; \left( \sum_{s=0}^\infty {1\over s!}
 {1\over {\cal T}_2^s} (-iD)^s ({\cal U}_2^2 {\bar \partial}_{\cal U})^s
\right)\
 {\cal A}({\cal U})\  + \  {\rm c.c.} \ .
\label{inst4}
\ee 
where here $D$ is the covariant derivative, which acting on a modular
form $\Phi_r$ of weight $-2r$ gives a form of weight $-2r +2$,
\be
D \Phi_r = \left( {i\over\pi}\partial_{\cal U} 
 - {r\over \pi {\cal U}_2}\right)
\Phi_r \ .
\ee
Some properties of covariant derivatives are summarized in the appendix.

The virtue of this last rewriting is that the $s$th operator in the
sum annihilates explicitly the first $s$ terms in the expansion
of the elliptic genus in powers of ${1\over {\cal U}_2}$.
From the general form of ${\cal A}$, eq. (\ref{exp}) we conclude
that 
 only the terms with $s\leq  2$ ($s\leq 1$) contribute in the
case of sixteen (eight) unbroken real supercharges.
 The modular-invariant
descendants of the genus, obtained  by applying the
 $s$th operator on ${\cal A}$,  
determine in fact  the corrections to  other dimension-eight
operators in the effective action. 
The full effective action can  be expressed in terms
of generalized holomorphic prepotentials, a result that we will
not develop further here.

\begin{figure}
\[
\psannotate{\psboxto(0cm;5cm){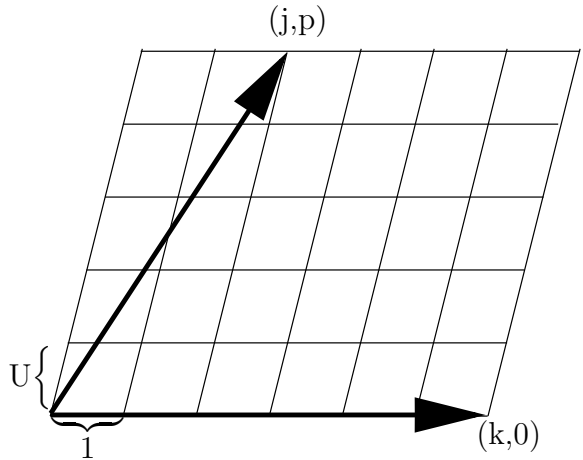}}
{
\at{6.5\pscm}{-.3\pscm}{(k,0)}
\at{3.5\pscm}{5.4\pscm}{(j,p)}
\at{-.2\pscm}{.5\pscm}{U$\left\{\vphantom{\vrule height.5cm }\right.$}
\at{.25\pscm}{.1\pscm}{$\underbrace{\hphantom{\vrule
width.95cm}}_{\displaystyle 1
}$}
}
\]

\caption{
Embedding of the lattice $\Gamma'$ (D1-brane) in the lattice $\Gamma$
(compactification torus).
\label{Lattic}}
\end{figure}

\section{D-instanton Interpretation}
\setcounter{equation}{0}

We would now like to understand the above result from the perspective
of type I string theory. 
The world-sheet instantons on the heterotic side map  to
D-brane instantons, that is Euclidean trajectories of D-strings
wrapping non-trivially around the compactification torus. 
A  Euclidean  trajectory
described  by eq. (\ref{wrap}) defines a sublattice ($\Gamma^\prime$)
 of the compactification
lattice ($\Gamma$).
 If ${\bf e}_{i=1,2}$ are the two vectors spanning $\Gamma$, then
$\Gamma^\prime$ is spanned by the vectors
${\bf e^\prime}_{i} = M_{ji} {\bf e_j}$ (figure \ref{Lattic}).
Under a change of basis for $\Gamma$ ($\Gamma^\prime$) the matrix
$M$ transforms by  left  (right) multiplication with
the appropriate elements  of SL(2,{\bf Z}). Using reparametrizations
of the  world-sheet we can thus bring the basis
 ${\bf e^\prime_i}$ into the canonical form, eq. (\ref{nondeg}),
 as described in
the previous section 
 (see also  figure \ref{Lattic}).

 Now the key remark is that on  the heterotic  world-sheet 
we have an  induced  complex structure and K{\"a}hler modulus,  
which for positive $p$ are given by
\be
{\cal U}= {j+pU\over k}\ \ \ {\rm and } \ \ \ 
{\cal T}= kp T \ .
\ee
For negative $p$'s,  describing  anti-instantons, we must take the
absolute value of $p$ and complex conjugate these  expressions.
One can check these facts by inspection of figure \ref{Lattic},
 or by computing
explicitly the
 pull-backs of the metric and
antisymmetric tensor field,
\be
\hat G_{\alpha\beta}=G_{\m\n}\partial_\alpha 
X^{\mu}\partial_\beta  X^{\nu}\,\,\,,\,\,\,
\hat B_{\alpha\beta}=B_{\m\n}\partial_\alpha
X^{\mu}\partial_\beta  X^{\nu} \ .
\label{31}
\ee
Notice that $N=kp$ is the total number of times the world-sheet
wraps around the compactification torus. 
In terms of induced moduli the instanton sum (\ref{inst})
 takes the form
\be
{\cal I}_{inst} \simeq -{2 V^{(10)}\over 2^{10}\pi^6}  \sum_{instantons}
      {1\over  {\cal T}_2}\;  e^{2\pi i {\cal T}}
        \;  {\cal A}({\cal U}) +  {\rm c.c.} \ .
\label{inst2}
\ee

\begin{figure}
\begin{center}
\leavevmode
\epsfbox{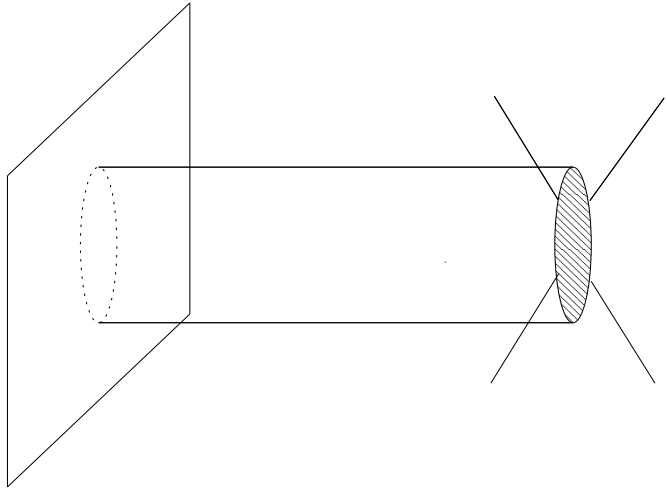}
\end{center}
\caption[]{A D1-brane instanton correction to $tr F^4$.}
\label{f5}\end{figure}

The various terms of this expression have a simple interpretation
on  the type I side. The action of a wrapped D-string is
 \cite{tasi}
\be
S_{\rm D-string}=
{1 \over 2\pi  \alpha'\lambda_I}\int d^2\sigma \sqrt{|{\rm det}
\hat G_I |}-{ i\over 2\pi \alpha'}\int \hat B_I
\label{30}
\ee
where $B_I$ is the type I 2-form coming from the RR sector.
Using the heterotic/type I map
\be
{\cal T}_2^{het} = {\cal T}_2^I/\lambda_I \ , \ \  B^{het} = B^I
\ee
and the fact that the world-sheet area of the D-string 
is $4\pi^2 {\cal T}_2^I$, we see that the exponential of this
Nambu-Goto action reproduces exactly the  exponential in the instanton
sum, eq. (\ref{inst2}). The inverse factor of the volume
comes from the integration of the longitudinal translation zero modes.
Finally the elliptic genus of the D-brane complex structure, 
should come from the  functional integration  over the
(second quantized) string fields in the instanton 
background. A typical  diagram contributing to the ${\cal F}^4$
coupling is shown in figure 3.
 For the purely holomorphic pieces of the elliptic
genus the result is topological, so it should be  expected to coincide
with the heterotic
 $\sigma$-model calculation of refs. \cite{Schellekens,Lerche,Windey,ellwit}.
Put differently, massive string modes and higher-order terms in
the effective D-string action are expected  to play no role in the 
calculation.
It is an  interesting and open problem to obtain  this result directly
on the type I side.

  The other interesting lesson from expression (\ref{inst}) concerns
the counting of distinct instanton solutions. The prescription in this case
is to include all supersymmetric
(holomorphic)  wrappings  modulo world-sheet
reparametrizations of the D-brane. One may conjecture that this
prescription stays valid for higher-dimensional branes, provided
one also mods out world-volume gauge symmetries when present.
This statement sounds obvious for world-sheet instantons on the
heterotic side, but is  non-trivial when considering for
example the solitonic five-brane.


\section{Concluding  Remarks}

 Perhaps the most interesting question  raised in this paper, is
the  calculation of the D-brane instanton contribution to the
effective action. 
Although the topological nature of this calculation makes it
plausible that the (leading)  answer should be proportional to the
elliptic genus, as suggested by the heterotic/type I duality, 
 it would be very interesting to see how this will
come about from explicit  string diagrams. 
This is  important, since it would open the way for
doing other semiclassical D-brane instanton calculations, particularly
in the background of the type I  D5-brane. This latter is a
heterotic zero-size instanton \cite{Witt}, for which
the field-theoretic calculation rules remain to be found.

 Another interesting check
would be the explicit evaluation of the higher-order perturbative
contributions. Depending on the world-sheet regularization, these
could appear through  corrections to lower-dimension operators,
as in the case of vacua with eight unbroken supercharges \cite{ABFPT}.
We believe that the presence of these terms is enforced by
supersymmetric Ward identities, and it would be interesting
to derive these in detail. Similar issues actually arose in the
study of D4-D0 brane scattering \cite{5d}, where the background geometry
seems to require a subleading two-loop open-string contribution.
Finally, we find  particularly intriguing the way in which string
theory regularizes what seems otherwise as a field-theoretic super
Yang-Mills expression.  It could be very interesting to contemplate
similarly the eleven-dimensional supergravity loop
 \cite{GGV}, whose regularization
may admit an analogous geometric interpretation.

\vskip 1cm

{\bf Acknowledgments}

  This research was partially supported by EEC grants CHRX-CT93-0340
and TMR-ERBFMRXCT96-0090. 
We  acknowledge the hospitality of the Newton Institute
(C.B., E.K., P.V.), of the high-energy theory group at Rutgers (C.B.),
of the DAMTP at Cambridge (P.V.), and of the CPTH at the Ecole
Polytechnique (E.K.,N.O.) during various stages of the work.
We also thank  E. D'Hoker, M. Green, J. Louis,
 M. de Roo, A. Tseytlin,   E. Verlinde and P. West 
for useful conversations.

\vskip 2cm
\newpage
\appendix
\section{Modular functions}

Holomorphic modular forms $\Phi_r(\tau)$ of weight $-2r$ are invariant under
$\tau\to \tau+1$ and transform as
\be
\Phi_r \to \tau^{2r}~ \Phi_r\;\;\;{\rm under} \;
\;\; \tau\to -{1\over \tau} \pe
\label{A1}\ee
The set of modular forms, relevant for our purposes, are the Eisenstein
series
\be
E_{2k} = - {(2k)!\over (2\pi i)^{2k} B_{2k}}\ G_{2k}\ ,
\label{A2}\ee
with  $B_{2k}$  the Bernouilli numbers  and
\be
G_{2k}(\tau) = \sum_{(m,n)\not= 0}\ (m\tau + n)^{-2k} \pe
\ee
for $k>1$.
For $k=1$ the Eisenstein series diverges. Its modular
invariant regularization,  denoted by a hat and used  in this paper,
is
\be
{\hat G}_2 (\t)  = \limit{lim}{s\to 0} \sum_{(m,n)\not= 0}\ (m\tau +
n)^{-2}
 \vert m\tau + n\vert ^{-s}
\pe \ee
The (hatted)  Eisenstein series are modular forms of weight $2k$.
The ring of holomorphic modular forms is generated by $E_4$ and $E_6$.
If we include (non-holomorphic) covariant derivatives (to be discussed
below)
then the generators of this ring are $\hat E_2$, $E_4$, $E_6$.

Expressed as power series in $q=\exp(2i\pi\tau)$,
 the first few of the Eisenstein
series
 are
\be
E_{2}(q)
=1-24\sum_{n=1}^{\infty}{n\, q^n\over 1-q^n}
\, \
\label{A3}
\ee
\be
E_{4}(q)
=1+240\sum_{n=1}^{\infty}{n^3q^n\over 1-q^n}
\, \
\label{A4}
\ee
\be
E_{6}(q)
=1-504\sum_{n=1}^{\infty}{n^5q^n\over 1-q^n}
\pe
\label{A5}\ee
The modified first Eisenstein series is
\be
 {\hat E}_2 =
 E_2-{3\over \pi\tau_2} \pe
\label{a6}\ee

The Dedekind function is
\be
\eta(\tau)=q^{1/24}\prod_{n=1}^{\infty} (1-q^n)
\ee
We can write the (weight 12) cusp form $\eta^{24}$ and the modular
invariant $j$-function in terms of $E_4$ and $E_6$
\be
\eta^{24}={1\over 2^6\cdot 3^3}\left[E_4^3-E_6^2\right]
\;\;\;,\;\;\;
j={E_4^3\over \eta^{24}}={1\over q}+744+\cdots
\label{a7}\ee

There is a (non-holomorphic) covariant derivative that maps modular
forms
of weight $-2r$ to forms of weight $-2r+2$:
\be
\Phi_{r-1}=\left({i\over \pi}\partial_{\t}-{r\over
\pi\tau_2}\right)\Phi_{r}=
-2\left(q\partial_q+{r\over 2\pi\tau_2}\right)\Phi_{r}
\equiv D\;\Phi_{r} \pe
\label{a9}\ee
The covariant derivative satisfies the Leibnitz rule:
\be
D~(\Phi_{r_1}\Phi_{r_2})= \Phi_{r_1}D\Phi_{r_2}+
(D\Phi_{r_1})\Phi_{r_2} \pe
\label{a8}\ee
Note that  a double
derivative on a weight $-2r$ form is
\be
D^2 \Phi_r\equiv\left({i\over \pi}\partial_{\t}-{r-1\over
\pi\tau_2}\right)\left({i\over \pi}\partial_{\t}-{r\over
\pi\tau_2}\right)
\Phi_r \pe
\label{a11}\ee

The following formulae allow the computation of the covariant
derivative
of any form:
\be
D\;\hat E_2={1\over 6}E_4-{1\over 6}\hat E_2^2
\;\;\;,\;\;\;
D\;E_4={2\over 3}E_6-{2\over 3}\hat E_2\;E_4
\label{a12}\ee
\be
D\;E_6=E_4^2-\hat E_2\;E_6 \pe
\label{a13}\ee

\end{document}